\documentclass[twocolumn,letterpaper,aps,prd,superscriptaddress,showpacs,nofootinbib,floatfix]{revtex4-1}
\usepackage{graphicx}
\begin{document}

\title{Search for exotic short-range interactions using paramagnetic insulators}

\author{P.-H.~Chu}
\email{pchu@lanl.gov}
\affiliation{Triangle Universities Nuclear Laboratory and Department
  of Physics, Duke University, Durham, North Carolina 27708, USA}
\affiliation{Los Alamos National Laboratory, Los Alamos, New Mexico 87545, USA }
\author{E.~Weisman}
\affiliation{Department of Physics, Indiana University, Bloomington IN
  47405\\ and IU Center for Exploration of Energy and Matter,
  Bloomington IN 47408} 
\author{C.-Y.~Liu}
\affiliation{Department of Physics, Indiana University, Bloomington IN
  47405\\ and IU Center for Exploration of Energy and Matter,
  Bloomington IN 47408} 
\author{J.~C.~Long}
\email{jcl@indiana.edu}
\affiliation{Department of Physics, Indiana University, Bloomington IN
  47405\\ and IU Center for Exploration of Energy and Matter,
  Bloomington IN 47408} 
\date{\today}

\begin{abstract}
We describe a proposed experimental search for exotic spin-coupled
interactions using a solid-state paramagnetic insulator.  The
experiment is sensitive to the net magnetization induced by the exotic
interaction between the unpaired insulator electrons with a dense,
non-magnetic mass in close proximity.  An existing experiment has been
used to set limits on the electric dipole moment of the electron by
probing the magnetization induced in a cryogenic gadolinium gallium
garnet sample on application of a strong electric field.  With suitable
additions, including a movable source mass, this experiment can be
used to explore ``monopole-dipole'' forces on polarized electrons with unique or
unprecedented sensitivity.  The solid-state, non-magnetic
construction, combined with the low-noise conditions and extremely sensitive
magnetometry available at cryogenic temperatures could lead to a
sensitivity over ten orders of magnitude greater than exiting limits
in the range below 1~mm. 
\end{abstract}

\maketitle

{\it Introduction.---}
Experimental searches for macroscopic forces beyond gravity and
electromagnetism have received a great deal of attention in the past
two decades. Present limits allow for unobserved forces several
million times stronger than gravity acting over distances 
of a few microns. Predictions of unobserved forces in this range
have arisen in several contexts, including attempts to describe
gravity and the other fundamental interactions 
in the same theoretical framework.  For comprehensive reviews,
see~\cite{Adelberger:2009zz,Antoniadis:2011zza,Jaeckel:2010ni}.  The
sub-millimeter 
range has been the subject of active theoretical investigation,
notably on account of the prediction of ``large'' extra 
dimensions at this scale which could explain the hierarchy
problem~\cite{ArkaniHamed:1998rs}.  The fact that the dark energy
density, of order (1~meV)$^{4}$, 
corresponds to a length scale of about 100~$\mu$m also encourages searches
for unobserved phenomena at this scale.  Many theories beyond the
Standard Model possess extended symmetries that, when broken at high
energy scales, give rise to light bosons with very weak couplings to
matter.  Examples include moduli~\cite{Dimopoulos:2003mw},
dilatons~\cite{Kaplan:2000hh}, and the axion--a light pseudoscalar
motivated by the strong CP problem of QCD~\cite{Peccei:1977hh}.  These
particles can generate weak, relatively long-range interactions between
samples of ordinary matter, including interactions that couple to
spin.  In a seminal paper~\cite{Moody:1984ba}, Moody and Wilczek derived
three possible interactions for the axion, and proposed searches
sensitive to the $T$-violating ``monopole-dipole'' interaction between polarized
and unpolarized test masses.

We propose an experimental search for exotic spin-coupled interactions
using a solid-state paramagnetic insulator as a detector.  The
candidate material is gadolinium gallium garnet 
(Gd$_{3}$Ga$_{5}$O$_{12}$, or GGG), which has been used as a
detector in an experimental search for the electric dipole moment (EDM) of the
electron~\cite{Kim:2011tn}. (See~\cite{Sushkov:2009a,Eckel:2012aw} for
realizations of other EDM materials.)  In that experiment, the signal
is an induced sample magnetization on application of a strong external
electric field; in our proposal the magnetization is induced by an
exotic monopole-dipole interaction with a
dense, non-magnetic mass brought into close proximity.  An
exotic field coupling to the electron spins of an atom or an ion with
spin $S=\sqrt{s(s+1)}$ in the solid sample leads to a net spin excess
in the sample.  At the location of a particular ion in the sample, the
spin excess ratio is given by~\cite{Kittel:1980b,Abragam:1978b} 
\begin{equation}
R=\frac{\displaystyle\sum_{m_{s}=-s}^{s} m_{s}e^{\frac{m_{s}u/\sqrt{s(s+1)}}{kT}}}{\displaystyle\sum_{m_{s}=-s}^{s} e^{\frac{m_{s}u/\sqrt{s(s+1)}}{kT}}}, 
\label{eq:r1}
\end{equation}
where $m_{s}$ is the magnetic quantum number, $u$ is the local exotic field
energy per ion, $T$ is the sample temperature and $k$ is Boltzmann's
constant.  In our proposal,
the energy shift predicted from the exotic coupling at the current
experimental 
limits is much smaller than the thermal energy.  However, the
cumulative effect from the large number of electrons in a macroscopic,
solid-state sample leads to a net spin alignment on the order of $10^{9}$
Bohr magnetons~cm$^{-3}$.  The resulting sample 
magnetization can be probed with great sensitivity using
sensors based on superconducting quantum interference device (SQUID)
technology. 

The idea of using solid-state materials with bound but unpaired electron
spins to probe exotic fields was first
proposed by Shapiro in 1968~\cite{Shapiro:1968a} in the context of EDM
searches.  Recently, other fundamental applications of these materials
have been proposed, including tests of Lorentz and $CPT$
invariance~\cite{Bluhm:1999ev}.  Experiments using EDM techniques, but
on a different class of solid-state materials, have been proposed to
search for cosmic axions~\cite{Budker:2013hfa}.  Many experiments have
been performed to 
search for spin-dependent macroscopic interactions using other
methods.  Examples include NMR-type experiments sensitive to
precession frequency shifts in various materials, including the
paramagnetic salt TbF$_{3}$~\cite{Ni:1999di}, Hg and Cs
comagnetometers~\cite{Youdin:1996dk}, polarized $^{129}$Xe and $^{131}$Xe
gas~\cite{Bulatowicz:2013hf}, and polarized $^{3}$He
gas~\cite{Vasilakis:2008yn,Chu:2012cf,Tullney:2013wqa}, in the presence of
polarized and unpolarized masses. 
Other experiments search for effects in torsion 
pendulums~\cite{Ritter:1993zz,Hammond:2007jm,Heckel:2008hw,Hoedl:2011zz},
neutron bound states in the Earth's gravitational 
field~\cite{Baessler:2006vm,Jenke:2014yel}, and longitudinal and transverse spin
relaxation of polarized neutrons and
$^{3}$He~\cite{Serebrov:2009ej,Pokotilovski:2009yf,Petukhov:2010dn,Fu:2011zzc,Petukhov:2011pr}. An  
overview can be found in~\cite{Antoniadis:2011zza}. Other parameters
being equal, the proposed solid-state technique affords an enhancement factor
on the order of the Avogadro number relative to experiments in dilute
vapor systems, though the latter are primarily sensitive to polarized nucleon
couplings.  With suitable control of systematic effects, the ultimate
sensitivity is more than 10 
orders of magnitude greater than current laboratory limits in the
range below 1~cm, and the technique is sensitive to
exotic interactions of electrons presently unconstrained by either laboratory
experiments or astrophysical observations. 

A study by Dobrescu and Mocioiu~\cite{Dobrescu:2006au} of the possible
interactions between non-relativistic fermions assuming only
rotational invariance revealed 
15 forms for the potential involving the fermion spins.  Nine of these
are spin-spin interactions, which would necessitate spin-polarized
test masses with low intrinsic magnetism~\cite{Ritter:1990a,Leslie:2014mua};
here we concentrate on monopole-dipole interactions 
between polarized and unpolarized objects.  In the
zero-momentum transfer limit, the possible 
interactions between a polarized electron and an unpolarized atom or
molecule of
atomic number $Z$ and mass $A$ are (in SI units, and adopting the
numbering scheme in~\cite{Dobrescu:2006au}): 
\begin{eqnarray}
V_{4+5} & = & \left(g_{A}^{e}\right)^{2}\frac{\hbar^{2}}{16\pi m_{e}c}Z\left[\hat{\sigma}\cdot(\vec{v}\times\hat{r})\right]\left(\frac{1}{\lambda r}+\frac{1}{r^{2}}\right)e^{-r/\lambda},
\nonumber \\
V_{9+10} & = & g_{P}^{e}g_{S}^{N}\frac{\hbar^{2}}{8\pi m_{e}}A(\hat{\sigma}\cdot\hat{r})\left(\frac{1}{\lambda r}+\frac{1}{r^{2}}\right)e^{-r/\lambda},
\nonumber \\
V_{12+13} & = & g_{A}^{e}g_{V}^{N}\frac{\hbar}{4\pi}A(\hat{\sigma}\cdot\vec{v})\left(\frac{1}{r}\right)e^{-r/\lambda}.
\label{eq:v1213}
\end{eqnarray}
Here $\vec{S}=\hbar\hat{\sigma}/2$ is the electron spin, $\hbar$ is
Planck's constant,
$\hat{r}=\vec{r}/r$ is a unit vector along the direction between the
electron and atom, $\vec{v}$ is their relative velocity, $c$ is the
speed of light in vacuum, $m_e$ is the electron mass, and $\lambda$ is the
interaction range.  The factors $g_{P}^{e}$ and $g_{A}^{e}$ are the
electron pseudoscalar and axial vector coupling constants, and
$g_{S}^{N}$ and $g_{V}^{N}$ are the nucleon scalar and vector
couplings.  The couplings in Eq.~\ref{eq:v1213} are not the most
general~\cite{Dobrescu:2006au}; we have included those for which the
proposed experiment will likely 
have the greatest discovery potential.  We note that $V_{9+10}$
can also proceed 
via a spin-1 interaction, in which case the coupling is
$g_{A}^{e}g_{V}^{N}$. $V_{4+5}$ can also proceed via a spin-0
interaction, in which case the coupling is $g_{S}^{e}g_{S}^{N}$ (and
the expression for $V_{4+5}$ above is scaled by $A/Z$).

\begin{figure}
\centering
\includegraphics[width=0.48\textwidth]{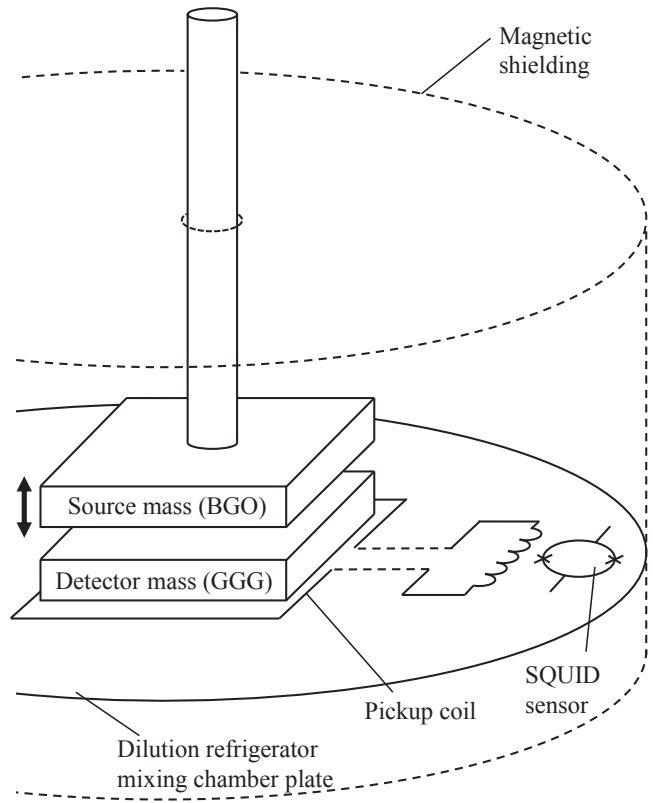}
\caption{Schematic of proposed experiment.  A dense, non-magnetic
  planar source mass is 
  modulated vertically above a paramagnetic detector mass,
  inducing a magnetization read out with the pickup coil and SQUID
  sensor.  The detector is kept at temperatures near 1~K.}
\label{fig:expt}
\end{figure}

The experiment is illustrated schematically in Fig.~\ref{fig:expt}.
It is based directly on the apparatus used in~\cite{Kim:2011tn}; many
parameters have been retained for the purposes of designing a practical
device.  A solid paramagnetic insulating sample or ``detector'' mass,
in the form of a block or disk, is mounted in the sample space of a
large dilution refrigerator.  A dense, unpolarized, non-magnetic insulator of
similar size and shape is brought into close proximity and serves as ``source''
mass.  The source--detector gap can be modulated (e.g., via
translation or rotation stages) from 0--1~cm with the resolution of a
few microns.  The non-magnetic design eliminates the need for
shielding between the test masses.  At closest approach, the source is
essentially in contact with the detector, permitting force searches
with potentially unprecedented sensitivity in the
range below 1~mm.  The detector magnetization induced by the exotic
interaction with the source is sensed by a pickup coil surrounding the
detector.  The coil is coupled to a sensitive SQUID magnetometer.  
 
For the detector material we assume GGG, as in~\cite{Kim:2011tn}. The use
of GGG was first proposed in~\cite{Lamoreaux:2001hb} and
realized in~\cite{Liu:2004rz}.  The material is chosen to maximize the
induced spin density.  GGG contains a high density of
Gd$^{3+}$ ions ($\approx 10^{22}$/cm$^{3}$), each of which has seven
electrons in the 4f shell: the shell is thus half-filled and all
electrons are unpaired~\cite{Geller:1978p}. This property leads to a
relatively strong magnetic response to external fields. For the
source mass we assume bismuth germanate (Bi$_{4}$Ge$_{3}$O$_{12}$,
or BGO), a high density, non-magnetic insulator~\cite{Tullney:2013wqa}.

{\it Experimental Sensitivity.---}
An approximate sensitivity and figure-of-merit can be derived using
a simplified geometry; we examine the case of the $V_{9+10}$
interaction, which is the most widely studied.  The interaction energy
between an unpolarized, 
flat--plate source mass parallel to a flat--plate paramagnetic
detector mass is given by:
\begin{eqnarray}
U_{9+10} & = & g_{P}^{e}g_{S}^{N}\frac{\hbar^{2}}{4m_{e}}A_{m}n_{m}n_{d}S_{d}a_{d}\nonumber\\
         &   &
        \times\lambda^{2}e^{-z(t)/\lambda}(1-e^{-t_{m}/\lambda})(1-e^{-t_{d}/\lambda}),
\label{eq:U910} 
\end{eqnarray}
where $n_{m}$ is the number density of molecules in the source mass
and $A_{m}$ is their mass number,
$n_{d}$ is the number density of polarized ions in the detector mass and
$S_{d}$ is the ion spin, $a_{d}$ is the detector area, $z(t)$ is the
source--detector gap, and $t_{m}$ and $t_{d}$ are the source and detector
thicknesses. This expression is exact for the case of a source of
infinite area, in which case the detector spins align toward the source,
in the direction normal to the detector plane.  For the practical
case of a finite
source (considered in detail below), there are edge corrections (which
are the dominant effect in the $V_{4+5}$ case; see Eq.~\ref{eq:v1213}).  The
proposed experiment is sensitive to changes in the induced field as
the source mass is 
modulated, which we assume to occur at some fixed frequency
$\omega$ significantly above the 1/f noise corner of the
SQUID.  The noise corner was about 0.1~Hz for the SQUID
in~\cite{Kim:2011tn}; we assume $\omega/2\pi=1$~Hz. From Fourier
analysis, 
the amplitude of the energy change per detector ion at $\omega$ is
\begin{equation}
\tilde{u}_{9+10}=A_{m}g_{P}^{e}g_{S}^{N}\frac{\hbar^{2}}{2m_{e}}n_{m}S_{d}\lambda I_{1}(z_{0}/\lambda)e^{-\bar{z}/\lambda}(1-e^{-t_{m}/\lambda}), 
\label{eq:u910}
\end{equation}
where $\tilde{u}_{9+10}=\tilde{U}_{9+10}/N_{d}$ and
$N_{d}=n_{d}a_{d}t_{d}$ is the 
number of detector spins.  In Eq.~\ref{eq:u910} we have taken the limit
$\lambda \gg t_{d}$, so that $\tilde{u}_{9+10}$ is a good approximation at the
location of any particular ion in the detector mass (or, equivalently,
in a layer of thickness $\approx \lambda$ at the top surface of a
thicker detector).  Here,
$I_{1}(z_{0}/\lambda)$ is the modified Bessel function, and we have
used $z(t)=\bar{z}+z_{0}\sin{\omega t}$ where
$\bar{z}=(z_{\mbox{\scriptsize max}}+z_{\mbox{\scriptsize min}})/2$ is
the average source--detector gap and $z_{0}=(z_{\mbox{\scriptsize
    max}}-z_{\mbox{\scriptsize min}})/2$ is the amplitude of the
source mass motion.

In the limit $\tilde{u}_{9+10}\ll kT$, the spin excess ratio
(Eq.~\ref{eq:r1}) can be approximated as $R=\tilde{u}_{9+10}/(3kT)$,
whereby the
total spin excess in the layer is $S_{d}N_{d}\tilde{u}_{9+10}/(3kT)$. 
The layer magnetization normal to the plane is
\begin{equation}
M = \frac{g\mu_{B}S_{d}n_{d}\tilde{u}_{9+10}}{3kT},
\label{eq:M}
\end{equation}
where $g = 2$ for electrons and $\mu_{B}$ is the Bohr magneton.  
As in~\cite{Kim:2011tn}, we assume the plane of the pickup coil to be
situated just below the bottom surface of the detector mass
(Fig.~\ref{fig:expt}).  The field of the magnetized slab of thickness
$\lambda$ just below the bottom is slowly-varying across the surface, with a
minimum at the center of $B_{z}\approx \mu_{0}M\lambda/\sqrt{a_{d}}$.
Assuming a coil area of $a_{d}$, a conservative estimate of the
induced flux
through the coil is thus $\Phi_{i}\approx\mu_{0}fM\lambda\sqrt{a_{d}}$, where
we include a suppression factor $f$ for sub-optimal coupling between
the detector and coil.    

In an experiment using a practical SQUID sensor, the pickup coil
connects to a built-in input coil on the sensor.  The changing flux
from the detector induces a current, which flows into the input coil and
produces a flux that couples inductively to the SQUID loop and is
transduced to a voltage.  The relationship between the magnetic flux
$\Phi_{i}$ picked up from the detector mass and the induced
flux $\Phi_{sq}$ in the SQUID loop is~\cite{Kim:2011tn}
\begin{equation}
\Phi_{sq} = \frac{M}{L_{p}+L_{i}} \Phi_{i} = \beta \Phi_{i},
\label{eq:beta}
\end{equation}
where $L_{p}$ and $L_{i}$ are the inductances of the pickup and input
coils, respectively, and $M$ is the mutual inductance between
the input coil and SQUID.  The factor $\beta$ is the coupling efficiency which
quantifies the flux reduction when $\Phi_{i}$ is delivered to the SQUID sensor. 

The sensitivity of the experiment is based on the expectation that
essentially all backgrounds can be suppressed below the intrinsic
noise of the SQUID sensor.  In an experiment limited by this noise
($\phi_{n}$, expressed in terms of magnetic flux per root bandwidth),
the signal-to-noise ratio is given by: 
\begin{equation}
SNR = \frac{\Phi_{sq}\sqrt{\tau}}{\phi_{n}},
\label{eq:snr}
\end{equation} 
where $\tau$ is the integration time.
The sensitivity is calculated by setting $SNR=1$ and
solving for $g_{P}g_{S}$. Combining Eqs.~\ref{eq:U910}
through~\ref{eq:snr}, the result is:
\begin{equation}
g_{P}^{e}g_{S}^{N}=\frac{4\mu_{B}m_{e}\mbox{u}}{\hbar^{2}}\frac{1}{\rho_{m}}\frac{1}{\lambda^{2}\epsilon}\frac{1}{\beta f}\frac{\phi_{n}}{\sqrt{\tau}}\frac{1}{\sqrt{a_{d}}}\left(\frac{1}{\chi_{d}}+D\right).
\label{eq:gg}
\end{equation}
Here we have used $n_{m}=\rho_{m}/(\mbox{u}A_{m})$, where $\rho_{m}$
is the mass density of the source and u is
the atomic mass unit.  The factor $\chi_{d}$ is the effective
susceptibility of the detector mass:
\begin{equation}
\chi_{d}=\frac{4\mu_{0}\mu_{B}^{2}n_{d}S_{d}^{2}}{3kT},
\label{eq:chi}
\end{equation}
and we have used $\chi_{d}\rightarrow\chi_{d}/(1+D\chi_{d})$ in
Eq.~\ref{eq:gg}, where $D$ is the demagnetization factor.  
The efficiency $\epsilon \equiv
I_{1}(z_{0}/\lambda)e^{-\bar{z}/\lambda}(1-e^{-t_{m}/\lambda})$ in
Eq.~\ref{eq:gg} is of order unity for an optimized vertical geometry.

\begin{figure}
\centering
\includegraphics[width=0.48\textwidth]{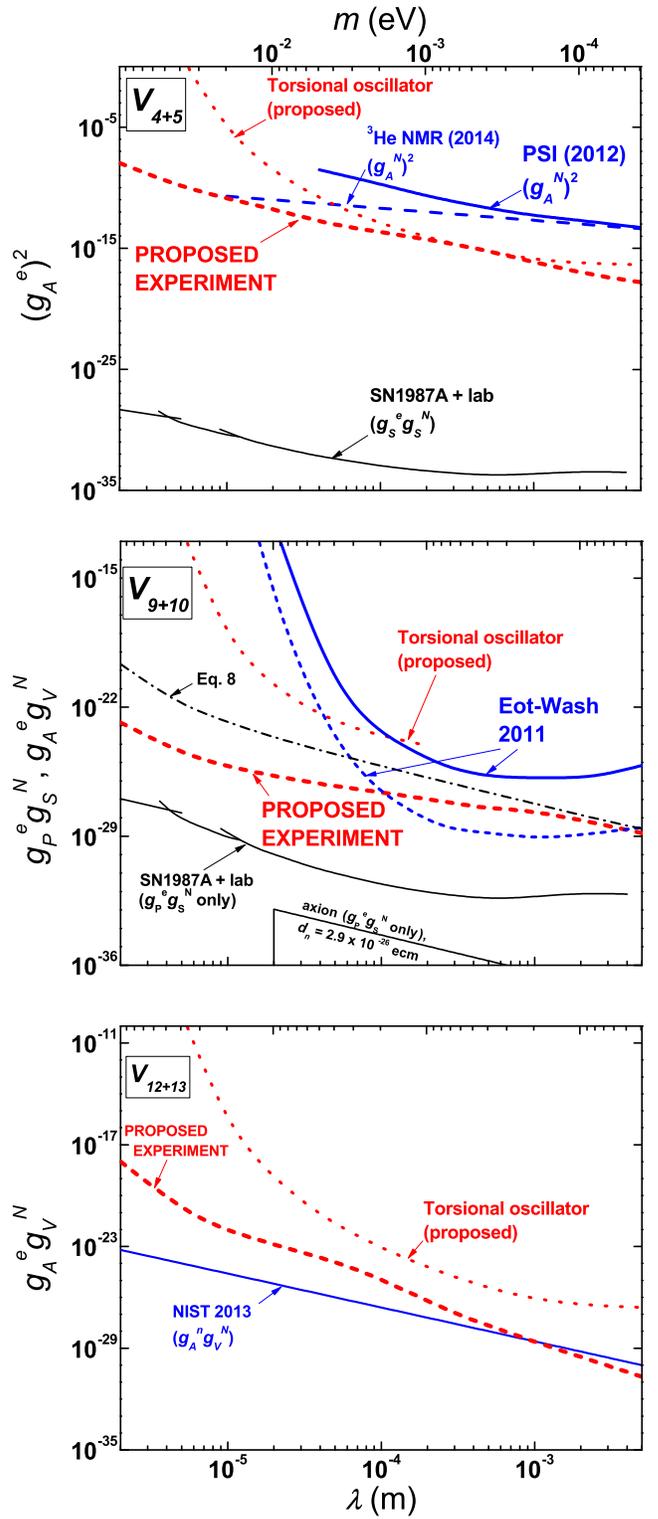}
\caption{(Color online) Projected sensitivity of proposed experiment to
  $V_{4+5}$, $V_{9+10}$, and $V_{12+13}$ interactions, in which couplings are
  plotted versus range
  $\lambda$ (lower axes) or mass of unobserved boson (upper axes).
  Bold solid (dashed) curves are current (preliminary) limits, fine
  solid curves are indirect limits, dotted curves are proposals,
  dot-dash curve is Eq.~\ref{eq:gg}. See text for explanations.}
\label{fig:limits}
\end{figure}

A plot of $g_{P}^{e}g_{S}^{N}$ vs $\lambda$ from Eq.~\ref{eq:gg} is shown in
Fig.~\ref{fig:limits}, using the parameters in
Tables~\ref{tab:parameters} and~\ref{tab:volume}. We assume the
source and detector can be brought into 
contact but also that these elements can only be made flat to
within a few microns, thus the minimum gap is set to 10~$\mu$m. For
each value of $\lambda$, the maximum separation 
$z_{\mbox{\scriptsize max}}$ is chosen to maximize $\epsilon$,
which occurs around $z_{\mbox{\scriptsize max}}=3\lambda$. The
efficiency is maximized for $t_{m}\rightarrow\infty$, however, since
very little is gained for
$t_{m}>\lambda$, we set $t_{m}=1$~cm, about twice the maximum range of
interest.  A suppression factor of $f=0.4$ has been
used, in accordance with~\cite{Kim:2011tn}. For a rectangular
prism of square cross-section $a_{d}$ magnetized parallel to the
thickness (here $\approx \lambda$), we have used the approximation
$D\approx\sqrt{a_{d}}/(\sqrt{a_{d}}+2\lambda)$~\cite{Sato:1989a}. 

From Eq.~\ref{eq:gg}, an approximate figure-of-merit for the
experiment is: 
\begin{equation}
FOM=\rho_{m}\epsilon\beta f\frac{\sqrt{\tau}}{\phi_{n}}\sqrt{a_{d}}\frac{\chi_{d}}{1+D\chi_{d}},
\label{eq:fom}
\end{equation} 
illustrating the importance of high $\rho_{m}$, and $\bar{z}\approx
z_{0}\approx\lambda$ (through $\epsilon$). Over the range of interest,
$D$ varies 
between 0.75 and 1.  Thus it is important that $\chi_{d}$ be at least of order
unity, but larger values will yield little improvement for the chosen
detector geometry. We note Eq.~\ref{eq:chi} suggests $\chi_{d}$ could
be improved by several orders of 
magnitude by operation at the 
sub-Kelvin temperatures available in dilution refrigerators. However,
as discussed in~\cite{Kim:2011tn}, the susceptibility of practical
paramagnetic insulators is well described by a Curie-Weiss relation of
the form $\chi=C/(T-T_{CW})$, where $|T_{CW}|$ represents an effective
minimum temperature assuming the operating temperature is lower.
Following~\cite{Kim:2011tn}, we use $T_{CW}=-2.1$~K.  We assume 
an operating temperature of 
1.0~K, the lowest temperature at which the susceptibility obeys the
Curie-Weiss law~\cite{Schiffer:1995a}.  This leads to an effective
temperature of 
$T=3.1$~K in Eq.~\ref{eq:chi} and an estimate of $\chi_{d}=0.53$
(Table~\ref{tab:parameters}), slightly below the value measured
in~\cite{Schiffer:1995a} for single-crystal GGG. Optimization of
other terms in Eq.~\ref{eq:fom} is discussed below.

\begin{table}[htb]
\caption{\label{tab:parameters} Parameters used in the sensitivity computation.}
\begin{ruledtabular}
\begin{tabular}{lr}
Parameter&Value\\
\hline
effective area of pickup coil, $a_{c}$& 12.75~cm$^{2}$\\
source mass density (BGO), $\rho_{m}$&7.13~g/cm$^{3}$\\
detector spin density, $n_{d}$&$10^{22}/$cm$^{3}$\\
detector spin/Gd ion, $s$&$7/2$\\
detector susceptibility, $\chi_{d}$ & $0.53$ (SI)\\
coupling efficiency, $\beta$ & $7.8\times 10^{-3}$\\
sensor intrinsic noise, $\phi_{n}$ &$3\mu\Phi_{0}/\sqrt{\mbox{Hz}}$\\
integration time, $\tau$ & $10^{6}$~s\\
\end{tabular}
\end{ruledtabular}
\end{table}

\begin{table}[htb]
\caption{\label{tab:volume}Dimensions of the detector and source
  masses used in the sensitivity calculations.} 
\begin{ruledtabular}
\begin{tabular}{lr}
Parameter&Value\\
\hline
detector width, $x_d$&3.00~cm\\
detector length, $y_d$&3.00~cm\\
detector thickness, $z_d$&0.76~cm\\
source mass width, $x_m$&3.00~cm\\
source mass length, $y_m$&3.00~cm\\
source mass thickness, $z_m$&1.00~cm\\
minimum gap, $z_{min}$& 10~$\mu$m\\
\end{tabular}
\end{ruledtabular}
\end{table}

For more accurate estimates of the sensitivity of a practical
experiment to each potential in Eq.~\ref{eq:v1213}, we perform numerical
calculations, in which all approximations used above are relaxed. The
theoretical interactions from each potential due to a finite-sized source are
computed at representative points in a hypothetical, practically-shaped
detector and used to generate a spin excess profile as a function
of the spatial coordinates.  This profile is then used in a finite
element (FEA) model to create a
map of the induced flux $\Phi_{i}$ in the region of the detector and
pickup coil.

\begin{figure}
\centering
\includegraphics[width=0.48\textwidth]{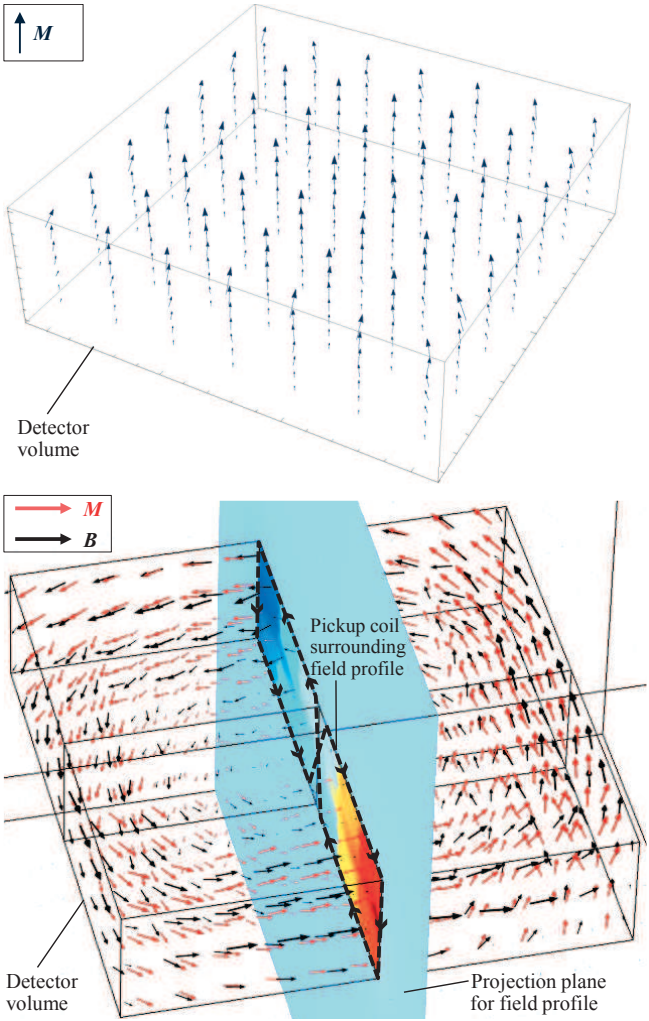}
\caption{(Color online){\it Top:} map of detector mass magnetization induced by
  source in $V_{9+10}$ interaction.  Vectors represent
  magnetization (each an average of one of 343
  subvolumes). {\it Bottom:} magnetization (light
  arrows) and magnetic field (black arrows) induced by source in
  $V_{4+5}$ interaction. Flux is calculated through cross section;
  dotted line shows placement of pickup coil for $V_{4+5}$ experiment. In both
  cases $\lambda = 5$~mm, and source (not shown) is slab of
  identical area centered above detector at instantaneous
  separation of 3~mm.}
\label{fig:M}
\end{figure}

The detector is broken up into
subvolumes and the potentials calculated by Monte Carlo integration
between the center point of 
each subvolume and the complete volume of the source mass.  The detector and
source dimensions (very similar to those used for the sample
in~\cite{Kim:2011tn}) are given in Table~\ref{tab:volume}.  The induced
spin orientation at each subvolume location is obtained by repeating
the calculation for many possible orientations and taking the
maximum.  The integration models the
modulation of the source mass (assumed sinusoidal) and records the
results at several values of the separation (that is, the source
phase) over a complete period.  A minimum source-sample gap of 10~$\mu$m
is used as in the analytical estimate.  The maxima are then converted
to magnetization vectors for each subvolume and phase using
Eq.~\ref{eq:r1} and the parameters in
Table~\ref{tab:parameters}. Fig.~\ref{fig:M}  
shows a detector magnetization map for the case of the $V_{9+10}$
interaction with $\lambda = 5$~mm at a particular source phase. 

Magnetization maps are then entered into an electromagnetic FEA model of
the sample generated with the COMSOL software
package~\cite{comsol}. The 
model interpolates between the points of the input data to generate
a complete magnetization profile within the detector volume and a
complete field profile in the interior and exterior.  An example field
profile is shown in Fig.~\ref{fig:M}.  

From these profiles, the
flux through the area of the pickup coils in the proposed
experiment is calculated.  Following~\cite{Kim:2011tn},
we assume a planar gradiometer design for the pickup coil for the
$V_{9+10}$ and $V_{12+13}$ interactions, to
reduce common-mode backgrounds; the coil area in
Table~\ref{tab:parameters} matches that in~\cite{Kim:2011tn}. For the
$V_{4+5}$ interaction, the coil takes the planar ``figure-8'' form
shown in Fig.~\ref{fig:M} for similar common-mode rejection, either
sandwiched between two halves of a split detector or wound through a
small vertical hole in the center.  For a
particular interaction in Eq.~\ref{eq:v1213} at a given range $\lambda$, the
experimental signal is calculated by taking the Fourier transform of
the flux though the gradiometer coil as a function of the source
phase, and scaling the result by the reduction factor in
Eq.~\ref{eq:beta}. For each value of
$\lambda$ investigated, the source mass amplitude is optimized for
maximum signal, resulting in amplitudes of order
$\lambda$.  Finally, the $SNR$ is obtained by dividing this result by
the sensor intrinsic noise (Table~\ref{tab:parameters}).  The coupling constants
in Eq.~\ref{eq:v1213} are then adjusted so that $SNR = 1$, resulting in
the sensitivity curves in Fig.~\ref{fig:limits}. 

The velocity-dependent interactions $V_{4+5}$ and $V_{12+13}$ are
{\it unconstrained} for the case of polarized electrons.  (We note
that, if $V_{4+5}$ is interpreted exclusively as an axial-vector
interaction, the
coupling $(g_{A}^{e})^{2}$ is 
tightly constrained by electron spin-spin
experiments~\cite{Ritter:1990a,Heckel:2013ina,Kotler:2015ura}.)  
For rough comparison,
the bold solid line in the $V_{4+5}$ plot is the limit on the corresponding
coupling for polarized nucleons from the experiment at the Paul
Scherrer Institute~\cite{Piegsa:2012th}; the long-dashed line is the preliminary
limit on the nucleon coupling from~\cite{Zhang:2014vdr}.  Similarly,
the bold solid 
line in the $V_{12+13}$ 
plot is the limit on the corresponding polarized nucleon coupling
derived from the neutron spin rotation experiment at NIST~\cite{Yan:2012wk}.

The best limit on the $V_{9+10}$ interaction for electrons derives from the
axionlike particle torsion pendulum in the Eot-Wash
group~\cite{Hoedl:2011zz}.  This limit 
is indicated by the bold line in the $V_{9+10}$ plot; the dotted line below
it is the projected thermal limit.  The sensitivity of the proposed
experiment ranges from two to over ten orders of magnitude greater in
the range of interest.  It is in good agreement with the analytical
estimate from Eq.~\ref{eq:gg} at large $\lambda$ where
$\lambda\approx t_{d}$, suggesting that $t_{d}$ better
estimates the depth to which the detector is significantly
magnetized.

The fine solid lines in the $V_{4+5}$ and $V_{9+10}$ plots are the
limits obtained from combining the constraints on $g_{S}^{e}$ or
$g_{P}^{e}$ from the 
inferred cooling rate of SN1987A with the constraints on $g_{S}^{N}$ from
spin-{\it independent} short-range fifth force
experiments~\cite{Raffelt:2012sp}.  These apply to spin-0
interactions and are thus for illustrative purposes only in the case of
$V_{4+5}$.  The $V_{9+10}$ limits are still more stringent than
the projections for
the direct search proposed, though the latter are more general and free
from uncertainties related to dense nuclear matter effects in stars.

The $V_{9+10}$ plot also shows the limit inferred from the
constraint on the electric dipole of the neutron
($d_{n}$)~\cite{Baker:2006ts}, which sets bounds on $g_{S}^{N}$ via the
QCD $\theta$-term and thus is restricted to case of the (spin-0) axion. 
For generic light scalars unrelated to the strong CP problem, the
limits from direct searches are more stringent than those inferred
from EDM bounds over the range of
interest~\cite{Mantry:2014zsa}. Thus, correlating 
observations in EDM and macroscopic experiments could help distinguish
axions from more generic light scalars.

The light dotted lines in each plot are the projected sensitivity of the
proposed experiment in~\cite{Leslie:2014mua}.  That proposal is sensitive
to polarized electron coupling and competes directly, but our
projections are stronger by about 1-10 orders of magnitude for all
interactions.  A proposal for a direct experiment sensitive to the
axion region in the $V_{9+10}$ plot is described
in~\cite{Arvanitaki:2014dfa}, though that experiment is sensitive to
polarized nucleon coupling. 

{\it Backgrounds.---} The analysis above assumes the sensitivity
of the experiment to be limited by the intrinsic noise of the SQUID
sensor.  Both GGG and BGO are good electrical insulators so Johnson
noise levels are expected to be very low, however, magnetic noise
due to dissipation in the test masses is another possible statistical
background.  The spectral density of magnetization
noise in the detector mass is given by~\cite{Eckel:2009a,Sushkov:2009a}:
\begin{equation}
M_{\omega}\approx\frac{1}{\mu_{0}}\sqrt{\frac{kT\mu^{\prime\prime}}{V\omega}},
\end{equation} 
where $V$ is the volume and $\mu^{\prime\prime}$ is the imaginary part
of the complex permeability.  While data on $\mu^{\prime\prime}$ for
the proposed GGG detector material are
not available, it is expected to be too small (and even smaller in
BGO, which has much lower susceptibility) for this noise to be
observable.  It was not observed, for example, in the experiment
in~\cite{Kim:2011tn}, results from which were consistent with SQUID noise after 5
days of integration time. 

Systematic effects are a greater concern.  
For example, the source can
acquire a magnetization $M_{m}=\chi_{m}B_{0}/\mu_{0}$ via the
interaction of its susceptibility $\chi_{m}$ with a stray field
$B_{0}$.  The changing flux through the pickup coil as the source
oscillates above it can mimic a signal.  BGO is diamagnetic with
$\chi_{m}=-1.9\times 10^{-5}$ at 
room temperature~\cite{Yamamoto:2003a}, however there is evidence that
it is weakly
paramagnetic in low fields.  Low-temperature data on $\chi_{m}$ of pure
BGO are not available, but the related compound
$\alpha$-Bi$_{2}$O$_{3}$ exhibits $\chi_{m}\approx 1.8\times 10^{-6}$
at 4~K in low fields~\cite{Kravchenko:2006a}.  We estimate the flux as 
$d\phi\approx (\partial B/\partial z)z_{0} a_{d}$, where $\partial B/\partial z$ is
the gradient above 
the center of a magnetized disk of area $a_{d}$ a distance $z_{d}$
above the coil (Table~\ref{tab:volume}), and $z_{0}$ is the largest
source amplitude. Requiring $d\phi < \phi_{n}/\sqrt{\tau}$ leads
to an upper limit on the allowed stray field of $B_{0}\approx
10^{-12}$~T.  Assuming typical stray laboratory 
fields on the order of $10^{-4}$~T, the required shielding factor is
$10^{8}$.  This is quite modest compared to the experiment
in~\cite{Kim:2011tn}, which used two layers of superconducting lead foils 
and three layers of mu-metal sheets wound on frames to attain a
measured shielding factor of $5\times 10^{11}$ (which we assume here).

Vibrations are potentially more troublesome.  Motion of the
detector-pickup coil assembly in phase with the source drive could
induce a signal in the presence of a stray gradient $\partial
B_{0}/\partial z$, given by $d\phi\approx (\partial B_{0}/\partial
z)\delta z_{d} a_{d}$, where $\delta z_{d}$ is the assembly vibration
amplitude.  This signal is common-mode; the common-mode rejection
ratio of the pickup coil used in~\cite{Kim:2011tn} was about $2\times
10^{2}$.  Dividing the estimate for $d\phi$ by this factor and again
demanding $d\phi < \phi_{n}/\sqrt{\tau}$ leads to a requirement on the
stray gradient of $\partial B_{0}/\partial z < 10^{-12}$~T/m per micron of
assembly vibration.  Assuming typical lab gradients of order $10^{-3}$~T/m, this
is within the shielding factor for vibrations less than about
100~$\mu$m.  If the vibrations cause the assembly to tilt in a stray
field of $10^{-12}$~T, the resulting flux signal estimate falls below the
noise as long as the vibrations are less than about 30~$\mu$m. 
These effects can
be studied by examining signals in the absence of a detector mass.  The
simple translating source mass assumed can be replaced by a rotor with several
segments of alternating density that pass over the detector at a
multiple of the rotary drive
frequency~\cite{Chen:2014oda,Arvanitaki:2014dfa}, thus decoupling the drive
frequency from the actual source modulation. 

{\it Conclusions.---} 
We have performed detailed calculations of the projected sensitivity
of an exotic interaction search with a paramagnetic insulating
detector at cryogenic temperatures.  The proposed technique affords
the possibility to
probe the interaction between macroscopic test masses in near contact
in a low-noise environment.  Our results indicate
either unique sensitivity to electron ``monopole--dipole''
interactions in the range 
below 1~mm, or improvements of more than ten orders of
magnitude over existing experiments.  The
statistical limits in Fig.~\ref{fig:limits} represent the
ultimate practical sensitivity of the experiment and are ambitious
long--term goals.  Results with
reduced but competitive sensitivity, likely limited by systematics,
are expected much sooner.  The proposed technique is based
largely on a proven design.  Our primary purpose has been
to show the potential sensitivity of that design, especially with the
parameters of the existing detection scheme.  Technical challenges associated
with source mass translation in the cryostat and the related systematic
backgrounds will certainly have to be addressed.  Possible
improvements include better SQUID coupling
efficiency ($\beta$ in Eq.~\ref{eq:fom}), though this is a subtle
optimization problem~\cite{Sushkov:2009a}.  Increasing the detector area $a_{d}$
(Eq.~\ref{eq:fom} scales as $\sqrt{a_{d}}$, given
the near saturation of the demagnetization factor) is another
possibility, though problems of test mass metrology and changes to
SQUID coupling efficiency will warrant careful study. 

{\it Acknowledgments.---}
The authors thank H. Gao, E. Smith, A. Holley, and
W. M. Snow for useful discussions.  This work is supported by National
Science Foundation Grants PHY-1207656, PHY-1306942, Duke University,
Los Alamos National Laboratory, and the Indiana University
Center for Spacetime Symmetries (IUCSS).

\bibliography{chu_exotics_in_paramagnets}

\end{document}